\documentclass[12pt]{article}
\usepackage{rotating}
\usepackage{epsfig,amssymb}
\input{epsf}

\def\laq{\raise 0.4 ex \hbox{$<$}\kern -0.8 em\lower 0.62 ex\hbox{$\sim$}}
\def\gaq{\raise 0.4 ex \hbox{$>$}\kern -0.7 em\lower 0.62 ex\hbox{$\sim$}}

\def\beq{\begin{equation}}
\def\eeq{\end{equation}}
\def\beqa{\begin{eqnarray}}
\def\eeqa{\end{eqnarray}}

 \def\frac#1#2{{\textstyle{{#1}\over {#2}}}}
 \def\lsim{\mathrel{\rlap{\lower4pt\hbox{\hskip1pt$\sim$}}
    \raise1pt\hbox{$<$}}} \def\gsim{\mathrel{\rlap{\lower4pt\hbox{\hskip1pt$\sim$}}
    \raise1pt\hbox{$>$}}}
\def\sqr#1#2{{\vcenter{\vbox{\hrule height.#2pt
         \hbox{\vrule width.#2pt height#1pt \kern#1pt
         \vrule width.#2pt}
         \hrule height.#2pt}}}}

\def\gappeq{\mathrel{\rlap {\raise.5ex\hbox{$>$}} {\lower.5ex\hbox{$\sim$}}}}

\def\lappeq{\mathrel{\rlap{\raise.5ex\hbox{$<$}}
{\lower.5ex\hbox{$\sim$}}}}

\begin{document}
\pagestyle{plain}

\begin{flushright}
June 2023
\end{flushright}
\vspace{15mm}

\begin{center}

{\Large\bf Seeding the vacuum with entropy: \\ the Chaplygin-like vacuum hypothesis}

\vspace*{1.0cm}

Orfeu Bertolami$^{*}$\\
\vspace*{0.5cm}
{Departamento de F\'\i sica e Astronomia, Faculdade de Ci\^encias, Universidade do Porto\\
Rua do Campo Alegre, 4169-007 Porto, Portugal}\\

\vspace*{2.0cm}
\end{center}

\begin{abstract}

\noindent
It is proposed that the vacuum admits two different phases as described by the Chaplygin equation of state or its generalised version:  a phase where the energy density behaves as if dominated by non-relativistic matter and a de Sitter phase. The particle production due to the expansion that takes place at the matter-like phase can generate entanglement entropy and provide, through interactions, the environment that turn gravitational quantum features into classical ones. In the ensued de Sitter phase, the cosmological constant can be suppressed by inflation.

\end{abstract}

\vfill
\noindent\underline{\hskip 140pt}\\[4pt]
$^{*}$ Also at Centro de F\'\i sica das Universidades do Minho e do Porto, Rua do Campo Alegre, 4169-007 Porto, Portugal. 
\noindent
{E-mail address: orfeu.bertolami@fc.up.pt}

\newpage
\section{Introduction}
\label{sec:intro}

The set of events and physical processes that gave origin to the Universe is one of the outstanding mysteries of contemporary science. The primordial onset is presumably intimately coupled with the quantum properties of gravity and the conceptual assumptions about the primordial fabric of space-time. The complex web of events arising from quantum gravity can be appreciated from the elaborate scenarios arising, for instance, from string theory and loop quantum gravity, just to mention two of the most studied quantum gravity proposals. However, it is not at all impossible that the very first beginning may lie in the realm of the physical processes that can be understood with the tools that we have been using in cosmology nowadays, namely Quantum Mechanics, General Relativity and Quantum Field Theory in curved space-time. Indeed, long ago, Tryon put forward the idea that the Universe could arise from a vacuum fluctuation, predicting that if so the emerging Universe would be homogeneous, isotropic, closed  and equally composed of matter and anti-matter \cite{Tryon}. Along similar lines, Vilenkin \cite{Vilenkin} proposed a model of a closed Universe that was created by quantum tunneling from nothing into a de Sitter space. After tunneling, the model would evolve generically like the inflationary model. In this proposal there is no big-bang singularity and no initial or boundary conditions are required. This scenario has common features with the no boundary quantum cosmology proposal of Hartle and Hawking to address the problem of initial conditions for inflation \cite{HH}. In fact, the discussion about the set of events that gave origin to the observable Universe most often include arguments of anthropic nature (see Ref. \cite{BT} for a thorough discussion) and more recently the so-called  swampland conjectures have been put forward to in order to select our 4-dimensional world from the vast set of possibilities within the string landscape \cite{Vafa,Douglas} (see Ref. \cite{Palti} for a review). 

Actually, the discussion of the principles that could give rise to General Relativity as an effective theory from a more fundamental underlying theory date back to proposals like the pregeometry programme, in which the metric field was composed of more fundamental matter fields \cite{Pregeo}, and proposals like the entropic gravity \cite{Verlinde}. However, effective field theory arguments have well-known limitations in what concern extracting gravity from a more fundamental theory (see, for instance, Ref. \cite{Burgess,Donoghue}).  

Despite the complexity of the discussion a few guidelines are available. For instance, it has been shown by Borde, Guth and Vilenkin \cite{BGV} that the history of an expanding Universe cannot be indefinitely scrutinised into the past. This means that as the volume of the Universe grows with time, inflation cannot be eternal and must have a beginning. A way out of the conclusion of the Borde-Guth-Vilenkin theorem is the idea of a cyclic Universe, which has been revived, for instance, in the context of M-theory \cite{ST}. However, in this scenario, the volume of the Universe is, on average, growing and the Borde-Guth-Vilenkin theorem can be applied. It follows that a cyclic Universe cannot be past-eternal. 
Furthermore, the Second Law of Thermodynamics requires that entropy should increase in each cycle of cosmic evolution. Thus, after an infinite number of cycles the Universe would have reached a state of thermal equilibrium, a state of maximum entropy, and thus the Universe would have a uniform temperature. Clearly, the Universe is not in such a state.

Besides these issues, one should have in mind the ubiquitous problem of the cosmological constant. In the present work we suggest that for describing the physics of the very early Universe prior to inflation it is interesting to consider a vacuum with two phases: an initial one where the Universe behaves as if dominated by non-relativistic matter and whose expansion gave the origin of the gravitational particle production generating entanglement entropy; and a second phase where the Universe is dominated by a cosmological constant. This evolution can be captured by a single equation of state, the Chaplygin equation of state or its generalised version. 

These assumptions allow for a particularly auspicious onset, as recently it has been pointed out that if the vacuum had a nonvanishing entropy then its energy density could be significantly suppressed after a period of inflation \cite{OB2021}. It was speculated that the origin of this entropy was some sort of dissipation mechanism like the one encountered, for instance, in warm inflation\footnote{Interestingly, the dissipation features of warm inflation are quite relevant for matching inflation with the so-called swampland conjectures (see, for instance, 
Refs. \cite{mkr,bkr,obs}.} \cite{Berera,Visinelli}. Of course, endowing the vacuum with entropy means that it has non-trivial properties and can undergo unexpected physical processes and transformations. The transition from the ``false" to the ``true" vacuum, the latter state corresponding to the lowest global energy  \cite{Coleman}, is an example of the possible transitions that can take place. However, here we propose a different form of vacuum transition, which will induce the above trnasition under special conditions. 

Indeed, the main point of our proposal is that the vacuum, whose pressure, $p_{V}$, must be negative, can be suitably described by the Chaplygin equation of state. This equation of state is a particular case of the generalised Chaplygin gas equation of state: 
\beq
p_{V} = - {A \over \rho_{V}^{\alpha}}~~,
\label{eq:eqstate}
\eeq
where $A$ is a positive constant, $0< \alpha \leq 1$ and $ \rho_{V}$ is the vacuum energy density. It is well known that this equation of state has many interesting features and is an insightful way to describe the properties of dark energy and dark matter in a unified manner \cite{Kamen,Bilic,BBS}. The support of the Chaplygin equation of state $(\alpha=1)$ is a Born-Infeld action that describes a brane \cite{Bilic} which can be parametrised through the action of a complex \cite{Bilic,BBS} or a real scalar field \cite{Kamen,BSST}.

Given that the vacuum is necessarily homogeneous and isotropic, its energy-momentum tensor is the one of a perfect fluid with vacuum energy and pressure related through the equation of state (\ref{eq:eqstate}).  In order to ensure the attractive nature of gravity at the very beginning, the energy conditions must be satisfied: the weak energy condition, $\rho_V \geq 0$; the strong energy condition,  $\rho_V + 3 p_V \geq 0$; the null energy condition, $\rho_V +  p_V \geq 0$. The dominant energy condition, $\rho_V -  p_V \geq 0$, is ensured by the weak energy condition once the equation of state, Eq. (\ref{eq:eqstate}) is considered. We assume that these broad features are sufficient to drive expansion in the context of the quantum gravity theory. 

Consistently with the above assumptions, the space-time metric should be, at least in average, the Robertson-Walker one, from which follows that the covariant conservation of the vacuum energy-momentum satisfies the equation: 
\beq
\dot{\rho}_V + 3 H (\rho_V + p_V)  = 0~~,
\label{eq:conservation}
\eeq
where the dot denotes the derivative with respect to the cosmic time, $H = \dot{a} /a$ is the expansion rate and $a(t)$ is the scale factor of the Robertson-Walker metric. 

Eq. (\ref{eq:conservation}) with eq. (\ref{eq:eqstate}) can be easily integrated \cite{Kamen,Bilic,BBS}:
\beq
\rho_V = \left(A + {B \over a^{3(1 + \alpha)}}\right)^{1 \over 1+ \alpha} = V_0  \left(1 + {1 \over a^{3(1 + \alpha)}}\right)^{1 \over 1+ \alpha}  ~~, 
\label{eq:rho_V}
\eeq
where $B$ is an integration constant, $V_0$ is a constant associated with the typical scale of inflation and the equality of constants $A$ and $B$ is imposed under the conditions that will be specified below.

Equation (\ref{eq:rho_V}) evinces the fact that, when the second term in eq. (\ref{eq:rho_V}) is the dominating one the vacuum energy behaves as a non-relativistic matter, which we refer to as B-phase;  on the other hand, when the first term in eq. (\ref{eq:rho_V}) is the dominant one, the Universe is dominated by a cosmological constant.    

In what follows, we shall argue that an interesting scenario arises from the assumption that the vacuum is described by eq.  (\ref{eq:rho_V}).

\section{The Chaplygin vacuum} \label{sec:ichaplygin-vacuum}

\subsection{The B-Phase}

Let us assume, just to fix ideas, that initially the $B/a^{3(1 + \alpha)}$ term in eq. (\ref{eq:rho_V}) is dominant. If we adopt the convention that $a_{Bf} = 1$, meaning that $a_{Bi} << a_{Bf}$, then the constants $A$ and $B$ can be identified, that is, $B=A$. Furthermore, the constant $A$ can be related to the typical energy scale of inflation, that is $2A= V_0^{(1+\alpha)}$, where $V_0=\Delta^4$, $\Delta \simeq 10^{-3} M$, and $M$ is the reduced Planck mass, $M=M_P/\sqrt{8\pi}$. The motivation for this simplification is to relate the end of the B-phase with the onset of inflation at the A-phase.  

The contribution of the B term implies in an expansion of the Universe that leads to a particle production according to well-establised principles of quantum field theory in curved space-time. The produced particles will have a collective entanglement entropy, which eventually becomes a thermodynamic entropy through interaction and decoherence. 

Indeed, assuming that initial space-time is a spatially flat Robertson-Walker Universe such that the scale factor in terms of the conformal time, $\eta$, is given by $a(\eta) = C + D~tanh(E \eta)$, where $C=(a_{Bi} +a_{Bf})/2$, $D=(a_{Bf}-a_{Bi})/2$,  $E$ is a constant, $a_{Bi}$ and $a_{Bf}$ are the initial and final values of the scale factor at the B-phase. As $a(\eta) \rightarrow C\pm D$ for $\eta \rightarrow \pm \infty$, it is shown that the particle production takes place  (see, for instance, Refs. \cite{BD,Ford} for extensive discussion and for a review). In fact, particle production can occur under more general conditions. We suggest that the entanglement entropy generated during this particle production is the vacuum entropy at the onset of the A-phase, which is the vacuum entropy that ensures the vacuum energy suppression by inflation as discussed in Ref.  \cite{OB2021}.

As the B-phase ends, it is followed by a period of inflation driven by constant A, the A-phase. In fact, an inflationary phase is helpful to ensure that some of the suitable theoretical conditions for particle creation are met under broad conditions, namely that $\it{in}$ and $\it{out}$ vacua correspond to flat space-time regions. In what concerns the $\it{out}$ state, inflation provides a natural solution as it drives space-time to be, in great approximation, flat. Inflation might also help to resolve the problem of the flatness of the $\it{in}$ state, as a long period of inflation erases any features of the preexisting initial conditions. However, as the B-phase precedes the A-phase, we have to assume that the $\it{in}$ state is asymptotically flat \cite{Ford}.

The proposed scenario is also helpful to hint at how the classical features of gravity and of inflation, in particular, arise so early on in the history of the Universe. Indeed, a  possible way to explain this puzzling feature is to assume that the very early Universe has undergone a process of decoherence due to the presence of interacting massive states that act as an environment and induce the emergence of the classical behaviour. A condition for decoherence and the emergence of the classical features can be written in terms of the scale factor, $a$:
\beq
a_{Bf} - a_{Bi} >> \left({m_0 \over M}\right)^{-1}~~,
\label{eq:classicality}
\eeq  
where $m_0$ is a typical mass scale at the B-phase. This  condition is fairly general and can be found in many models of decoherence \cite{Decoherence}, for instance, in the context of quantum cosmology (see eg. Ref. \cite{OBM95}).

\subsection{The A-Phase}

Let us now focus on the inflationary process as discussed in Ref. \cite{OB2021}. Following the terminology proposed by Coleman and co-workers \cite{Coleman}, we call the old or ``false" vacuum energy, $\rho_F$, the energy density at the ground state of the inflaton before inflation and ``true" vacuum energy, $\rho_T$, the ground state energy density after inflation.  The main point of Ref. \cite{OB2021} is that if the vacuum has an intrinsic entropy, then inflation can suppress significantly $\rho_T$. The details can be found in Ref. \cite{OB2021} and consist in starting from the Gibbs-Duhem equation to obtain the relationship between pressure, $p$, temperature, $T$, entropy, $S$ and volume, $V$ for the vacuum,
\beq
{dp \over dT} = {S \over V}~~ 
\label{eq:GDVequ}
\eeq
and use the Bekenstein entropy bound \cite{Bekenstein} to relate the entropy with the energy, $E$, and the length scale of the system, $R$:
\beq
S  \leq 2 \pi R E ~~,
\label{eq:Bekenstein}
\eeq
where we have set $\hbar=c=k=1$ ($k$ being Boltzmann's constant). Saturating the bound and assuming that the length scale corresponds to the horizon distance; hence Eq. (\ref{eq:GDVequ}) can be written as 
\beq
{dp \over dT}  =  2 \pi  R_{Hor}  \rho~~.
\label{eq:BekensVequ}
\eeq

Assuming that inflation is driven by a scalar field at a typical energy scale, $V_0$, thus the expansion rate is given approximately by $H_I \simeq \sqrt{V_0}/M$ and the scale factor evolves as $a(t)=a_{Ai}e^{H_It}$, where $a_{Ai}=a_{Bf}$ is the scale factor at the onset of inflation. It is shown that after integration one gets for relationship between the ``true" and ``false" vacuum energy densities \cite{OB2021}:
\beq
{p_T \over p_F} = e^{-2 \pi  I} ~~.
\label{eq:Thermodynr1}
\eeq
In order to estimate the integral $I$, it is assumed that the vacuum is an ordered and extremely hot state,, that is, $T < -M$. Then, through a suitable choice for the  integration measure, it follows \cite{OB2021}
\beq
I > {a_{Af} M  \over a_{Ai} H_I}~~,
\label{eq:I}
\eeq
where the index $f$ refers to the end of inflation. 

Therefore, writing the vacuum energy as  $\rho_F \simeq V_0$, then for $65$ e-foldings of inflation (see Ref. \cite{Olive} for a general discussion), $a_{Af}=a_{Ai} e^{65}$, leading to 
\beq
\rho_T \simeq e^{(-2 \pi { e^{65} M \over  H_I})} V_0~~,
\label{eq:Thermodynr2}
\eeq
which is a significant suppression, ${\cal O}(10^{-3 \times 10^{34}})$, for typical inflation values, as specified above, $H_I \simeq V_0^{1/2}/M$, $V_0=\Delta^4$, and $(\Delta/M) \simeq 10^{-3}$ \cite{OB2021}.

This is a surprising result and it has some interesting implications. To start with, the fine tuning issue is gone. Second, dark energy required to account for late accelerated expansion of the Universe cannot be due to a residual of the very high energy vacuum. We stress that the suppression is achieved thanks to the relation between the entropy and energy density and the former is related to the entanglement entropy due to particle creation that took place at the B-phase expansion.  This is the main point of the scenario proposed in this work. Notice that in here the origin of the entropy differs from the one speculated in Ref. \cite{OB2021}.  

Of course, the mechanism outlined above does not apply, as discussed in Ref. \cite{OB2021}, for cosmological phase transitions after inflation. Indeed, it is understood that the cosmological constant acquires contributions at least from the electroweak phase transition and from the quark-gluon phase transition. As pointed out in Ref. \cite{OB2021}, the generic thermodynamic features, such as the exponential drop of the pressure arising from the Clausius-Clapeyron equation for the change of the critical points and in the Kelvin-Helmholtz equation for the pressure change due to capillaries in fluids, might be present in second order phase transitions as described by the Cahn-Hilliard model \cite{Cahn-Hilliard}. Even though the required features of the mechanism outlined above do not seem to be ruled out by general considerations about cosmological phase transitions \cite{Kibble,Linde3,Ginsparg}, they clearly demand a detailed analysis to assess its feasibility for phase transitions after inflation.

\section{Discussion and Final Remarks}

The conceptual understanding of the vacuum has evolved throughout the history of physics. In classical physics and in quantum mechanics the vacuum is no more than the lowest energy state being devoid of any particular feature. This has changed with quantum field theory.  Decades of effort have shown that the vacuum manifests itself through the Casimir forces, it does gravitate, it can give origin to thermal radiation, undergo phase transitions and give rise to topological defects very much like in material systems. It is known that when the vacuum changes its properties it can lead to the spontaneous breaking of symmetries, which is a crucial ingredient in the Standard Model of Fundamental Interactions of Nature. This leads, as is well known, to the embarrassing fine tuning problem of the cosmological constant. In Ref. \cite{OB2021} it was proposed that this problem could be partially addressed if the vacuum had an entropy and through a period of inflation. In that work, a speculative discussion about the origin of this entropy in the context of warm inflationary models was advanced. In the present work, it is argued that the needed entropy can arise from the generation of particles produced from curved space quantum field effects at the very early expansion of space-time. This suggests that the vacuum could primordially have two phases as described by the Chaplygin gas or its generalisation.  
It is a humble suggestion, but it does not seem to contradict any of the fundamental principles and properties required to setup a consistent quantum gravity theory. In fact, given that, in this proposal, the A-phase of the vacuum is fundamentally attached to inflation, it suggests the need for criteria similar in spirit to the so-called swampland conjectures 
\cite{OPSV,garg-krishnan,Andriot} proposed in the context of the landscape of string theory in order to obtain consistent effective models from the fundamental quantum gravity theory. 

In summary, our proposal contains two main ideas. First, based on quantum field theory in curved space-time, in the B-phase, expansion itself can seed the vacuum of the subsequent A-phase with entropy and provide the environment to decohere the quantum features of the very early Universe physics and explain the early emergence of classical gravity. Afterwards, in the A-phase, the vacuum is dominated by a cosmological constant that leads to an inflationary phase which, besides solving all initial conditions problems, gives rise to the primordial energy density fluctuations and also suppresses the vacuum energy density.  

It is clear that the scenario proposed here strongly reinforces the paradigmatic status of inflation. Inflation is an absolutely necessary step to explain the main features of the observed Universe. Moreover, it provides evidence that the vacuum is not a passive and immutable entity, but instead, as suggested long ago, it evolves throughout the history of the Universe \cite{Bronstein,Bertolami86,Ozer}. 

From the observational point of view, broadly speaking, it follows from the assumptions considered here that the cosmological recent observed accelerated expansion of the Universe is not due to a residue of the cosmological constant and, in principle, the current equation of state of the Universe, $w \equiv {p \over \rho} \neq -1$. There is no shortage of candidates to play the role of dark energy (see Ref. \cite{Copeland} for a review). Obviously, nothing prevents the Claplygin gas model or its generalisation is, in a new avatar, once again a putative candidate as it allows for an elegant unification of dark matter and dark energy \cite{Kamen,BSST} provided it can suitably circumvent issues related to structure formation (see Ref. \cite{BBS_04} and refs. therein). 

To conclude,we can state that our suggestion allows for the thrilling possibility that the cosmological constant problem and the origin of the Universe can be addressed without a full understanding of the fundamental quantum gravity theory. In fact, the discussion in the context of string theory provides an example of how the search for consistent effective theories can help in selecting, within the huge set of vacua of the fundamental theory, a vacuum that is consistent with the observable Universe. It would be quite interesting that criteria along these lines could indicate that the vacuum should have more than one phase as described by the Chaplygin equation of state or its generalised version.   


\vspace{0.5cm}

{\bf Acknowledgments}

\noindent
The author acknowledges the partial support from Funda\c{c}\~ao para 
a Ci\^encia e a Tecnologia (Portugal) through the research project
CERN/FIS-PAR/0027/2021.



\bibliographystyle{unstr}

\end{document}